\documentclass[a4paper,pre,twocolumn,showpacs,amsmath,amssymb,floatfix,superscriptaddress]{revtex4}

\usepackage{dcolumn}   
\usepackage{bm}        
\usepackage{graphicx}  

\def\bea{\begin{eqnarray}}
\def\eea{\end{eqnarray}}
\def\be{\begin{equation}}
\def\ee{\end{equation}}
\def\bes{\begin{equation*}}
\def\ees{\end{equation*}}
\def\bi{\begin{itemize}}
\def\ei{\end{itemize}}

\def\etal{{\it et al.}}
\def\ie{{\it i.e.}}

\def\d{\mbox {d}}

\begin{document}

\title{Slow dynamics and precursors of the glass transition in
  granular fluids} \author{Iraj Gholami}\thanks{Author to whom
  correspondence should be
  addressed. iraj.gholami@theorie.physik.uni-goettingen.de}
\affiliation{Georg-August-Universit\"at G\"ottingen, Institut f\"ur
  Theoretische Physik, Friedrich-Hund-Platz 1, D-37077 G\"ottingen,
  Germany} \author{Andrea
  Fiege}
\affiliation{Max-Planck-Institut f\"ur Dynamik und Selbstorganisation,
  Bunsenstr. 10, D-37073 G\"ottingen, Germany} \affiliation{Georg-August-Universit\"at G\"ottingen, Institut f\"ur
  Theoretische Physik, Friedrich-Hund-Platz 1, D-37077 G\"ottingen,
  Germany}  \author{Annette
  Zippelius}
\affiliation{Georg-August-Universit\"at G\"ottingen, Institut f\"ur
  Theoretische Physik, Friedrich-Hund-Platz 1, D-37077 G\"ottingen,
  Germany} \affiliation{Max-Planck-Institut f\"ur Dynamik und
  Selbstorganisation, Bunsenstr. 10, D-37073 G\"ottingen, Germany}

\begin{abstract}
  
  We use event driven simulations to analyze glassy dynamics as a
  function of density and energy dissipation in a two-dimensional
  bidisperse granular fluid under stationary conditions. Clear
  signatures of a glass transition are identified, such as an increase
  of relaxation times over several orders of magnitude. As the
  inelasticity is increased, the glass transition is shifted to higher
  densities and the precursors of the transition become less and less
  pronounced -- in agreement with a recent mode-coupling theory. We
  analyze the long-time tails of the velocity autocorrelation and
  discuss its consequences for the nonexistence of the diffusion
  constant in two dimensions.
\end{abstract}



\maketitle


\section{Introduction}

The physics of nonequilibrium systems in general involves dissipation
and energy injection. These properties distinguish nonequilibrium but
stationary systems from those in thermodynamic equilibrium and render
the application of standard thermodynamics and statistical mechanics
impossible such that one has to develop new methods to treat
nonequilibrium systems or resort to experimental and computational
methods. Due to their inherent dissipative nature, driven granular
fluids serve as model systems for nonequilibrium phenomena and in
particular phase transitions in macroscopic systems out of
equilibrium. Questions addressed include the crystallization of
granular fluids \cite{urbach-2005,reis-2006}, the jamming transition
\cite{Liu-98} and the glass transition \cite{kranz-2010}.

Here we focus on a two-dimensional driven granular fluid, displaying
structural arrest as the density is increased. Most experimental work
in the field does indeed use a two-dimensional setup, allowing for
extensive particle tracking with high speed and resolution
cameras. Several driving mechanisms have been studied, such as vertical
vibrations \cite{urbach-2005,reis-2006,reis-2007} and air-tables
\cite{oger-1996,abate-2006,abate-07}. Since we are interested in
structural arrest, the most relevant experiments are those of Abate
and Durian \cite{abate-2006,abate-07}, who showed that the approach to
jamming in a granular medium resembles the glass transition in a
molecular or colloidal fluid: The mean square displacement develops a
plateau and the dynamics becomes strongly heterogeneous.

Recently a mode-coupling theory \cite{kranz-2010} has been set up for
a driven granular fluid in the stationary state. A glass transition
was located below random close packing which is qualitatively similar,
but quantitatively different from the corresponding transition for a
molecular fluid. The coherent scattering function was shown to display
a plateau -- provided the density is sufficiently close to the glass
transition. The timescale for relaxation from the plateau ($\alpha$-relaxation) diverges at the glass transition. The transition density
as well as the dynamics depend on the degree of inelasticity,
parametrized by the coefficient of restitution.

In this paper we investigate the dynamics of a driven two-dimensional
granular fluid close to the glass transition, mimicking an air-table
experiment. We use event driven simulations to compute the mean square
displacement, the velocity autocorrelation, and the incoherent scattering
function.

\section{Model}
Dense systems of monodisperse, elastic hard disks are known to undergo a
phase transition to a crystalline state \cite{alder-1962}. The only state
variable is the packing fraction $\eta$, defined by the area covered by
particles divided by the area of the whole system. The phase transition
occurs for packing fractions between $\eta_f=0.691$ and $\eta_s = 0.7163$,
where packing fractions up to $\eta_f$ are completely in the fluid phase and
packing fractions larger than $\eta_s$ are completely in the solid phase.
Packing fractions between $\eta_f$ and $\eta_s$ fall into the coexistence
area. An overview over simulations performed to determine the phase
transition is given in \cite{mitus-1997}. Crystallization can be
avoided in binary mixtures of particles with sufficiently different
sizes. Whereas bidisperse mixtures of elastic hard spheres form
stable crystalline structures for a size ratio smaller than $1.2$
\cite{speedy-1998}, larger size ratios suffice to avoid
crystallization. We expect similar behaviour for a granular fluid and
choose a bidisperse symmetric mixture of hard disks. The ratio of the
radius of the big particles, $R_b$, to the radius of the small
particles, $R_s$, is given by $R_b/R_s=1.43$.

\subsection{Collision rules}

We restrict our analysis to smooth particles without rotational degree
of freedom. Particles' positions and velocities are denoted by
$\{\mathbf{r_i},\mathbf{v_i}\}_{i=1}^N$. The dissipation arises solely
from incomplete normal restitution with a constant coefficient of
restitution $\varepsilon$: The change of the relative velocity
$\mathbf{g} := \mathbf{v_1} - \mathbf{v_2}$ in the direction of the
normal vector $\mathbf{n} :=
(\mathbf{r_1}-\mathbf{r_2})/|\mathbf{r_1}-\mathbf{r_2}|$ is given by
\be
(\mathbf{g} \cdot \mathbf{n})' = -\varepsilon (\mathbf{g} \cdot \mathbf{n}).
\ee
Here primed and unprimed quantities indicate postcollisional and
precollisional velocities of the two colliding particles. The
postcollisional velocities of the two colliding disks are expressed in
terms of the precollisional velocities according to
\bea
m_1\mathbf{v'_1}&=&m_1\mathbf{v_1}-\bm{\delta},\\
m_2\mathbf{v'_2}&=&m_2
\mathbf{v_2}+\bm{\delta},
\eea
where
\bea
\bm{\delta}=\frac{m_1m_2}{m_1+m_2}\frac{1+\varepsilon}{2}(\mathbf{g}\cdot
\mathbf{n})\mathbf{n}.
\eea
Here $m_i$ denotes the mass of particle $i$ and we assume constant
mass density for all particles such that the mass ratio of the big to
the small particles is given by $(R_b/R_s)^2$.

\subsection{Driving}
Due to inelastic collisions, energy is dissipated and we have to feed
energy into the system in order to obtain a stationary state. One of
the simplest driving mechanisms \cite{williams-1996} is to kick a
given particle, say, particle $i$, instantaneously at time $t$
according to
\be
\mathbf{p}_i'(t)=m_i\mathbf{v}_i'(t)=\mathbf{p}_i(t)+p_{\text{Dr}}\bm{\xi}_i(t).
\ee
Here $p_{\text{Dr}}$ is the driving amplitude and $\bm\xi_i(t)$ indicates
the direction of the driving which is chosen randomly with $\langle
\xi_i^{\alpha}(t)\xi_j^{\beta}(t')\rangle =
\delta_{ij}\delta_{\alpha\beta}\delta(t-t')$. The Cartesian components
$\xi_i^{\alpha}, \alpha=x,y$, are distributed according to a Gaussian
distribution with zero mean.

In practice, we implement the stochastic driving process by kicking the
particles randomly with frequency $f_{Dr}$. In order to conserve
momentum locally, we apply kicks of equal magnitude and opposite
direction to pairs of neighbouring particles
\cite{espanol-2005}. Thereby momentum is conserved locally.

\subsection{Implementation}
As long as the system is not jammed, the
collisions of hard particles are instantaneous, so that event driven
simulations can be applied. Lasting contacts, that we expect to occur
in a jammed state, cannot be accounted for by event driven
simulations, so that we are restricted to densities below the jamming
point. The glass transition in a driven granular fluid is expected to
occur at densities below jamming, so that the restriction is actually
less severe.

We use an efficient standard event driven algorithm \cite{lubachevsky} that enables us to simulate large systems for long times. Inelastically colliding hard particles are known to undergo the so-called \emph{inelastic collapse}. We circumvent the inelastic collapse with the same method used in \cite{fiege}.

We consider a system of $N=10000$ particles in an area $A=L^2$, where $L$ denotes the length of system. The mixture
is chosen symmetric, so that small and big particles appear in equal
numbers. Initial velocities are drawn from a Gaussian distribution.
To analyze the slow dynamics of the system as the glass transition is
approached, we simulate a range of coefficients of restitution, $0.5
<\varepsilon < 0.95$ and area fractions $0.1 \leq \eta \leq 0.81$.
For every parameter set $(\eta, \varepsilon)$ we performed an
event driven molecular dynamic simulation with approximately $10^7$
collisions per particle. 

Several initial configurations with $\eta=0.84$
were taken from a time driven molecular dynamics simulation of soft
disks \cite{Stark-2011}, that suffer dissipation and finally come to
rest. By expanding the system we change the packing fraction, allowing
us to investigate systems of hard disks up to $\eta=0.81$.
To be sure that our simulations do not depend on the initial
configurations (\ie, the initial values of particles' velocities and
positions), we have repeated the simulation for each parameter set, using
several different initial configurations.

In order to investigate the long-time behaviour of the mean square
displacement and the velocity autocorrelation function it was
necessary to simulate larger systems with $N=4\cdot 10^6$. This was done for
$\varepsilon = 0.7$, $0.8$, $0.9$ and $\eta = 0.1 \ldots 0.78$.

\section{Results}

Our main concern is the slow dynamics in a dense granular fluid and in
particular the precursors of the glass transition. Our analysis will
be based on the time dependent van Hove correlation function,
the mean square displacemnet (MSD) and the velocity autocorrelation
function (VACF) in the {\it stationary} state. Hence we briefly regress to
discuss the relaxation towards stationarity.

\subsection{Stationary state}

To maintain a stationary state, the dissipated energy due to inelastic
collisions must be compensated by a comparable driving energy fed into
the system. In a two-dimensional monodisperse system, we define the
granular temperature as the average kinetic energy 
\bea
T:=\frac{1}{N}\sum_{i=1}^N \frac{m}{2} \mathbf{v}_i^2.
\eea 
A stationary state can be achieved by balancing dissipation and driving
\cite{annette-PA-2006}
\bea 0=\frac{\d}{\d t} T &=& \frac{\d T}{\d t}\vert_{\tiny\mbox{Drive}} +
\frac{\d T}{\d t}\vert_{\tiny\mbox{Inelastic}} \nonumber\\
&\approxeq& f_{Dr} \frac{p^2_{Dr}}{2m} - \omega_E \frac{1-\varepsilon^2}{4} T.
\label{d-temp}
\eea
The collisional loss is estimated by the energy loss in a single collision,
$\frac{1-\varepsilon^2}{4} T$, multiplied by the Enskog collision frequency
$\omega_E$:
\be
\omega_E=2\pi g(2a) (2a)^{d-1}\frac{N}{A} \sqrt{\frac{T(t)}{\pi m}}.
\ee
Here $g(2a)$ is the pair correlation at contact.

In the following we are going to discuss correlations, such as the
MSD or the VACF as a
function of density and inelasticity $\varepsilon$. We are particularly 
interested in the approach to the glass transition as the inelasticity
is varied. To isolate effects of inelasticity and packing fraction, we
have to keep the temperature constant (without loss of generality we
take $T\sim 1$). This can be achieved approximately by adjusting the
driving (see Eq. \ref{d-temp}) according to
\be
\Longrightarrow p_{Dr} = \sqrt{\frac{1-\varepsilon^2}{2}\frac{m
\omega_E}{f_{Dr}}}. 
\label{d-mom}
\ee In order to sustain a stationary state we choose the driving
frequency of the order of the collision frequency as suggested by
Bizon \etal~\cite{bizon}. Note that the driving amplitude scales with
the coefficient of restitution, such that we are able to take the
elastic limit, $\varepsilon \to 1$.

\begin{figure}[t]
  \begin{center}

    \includegraphics[width=\columnwidth]{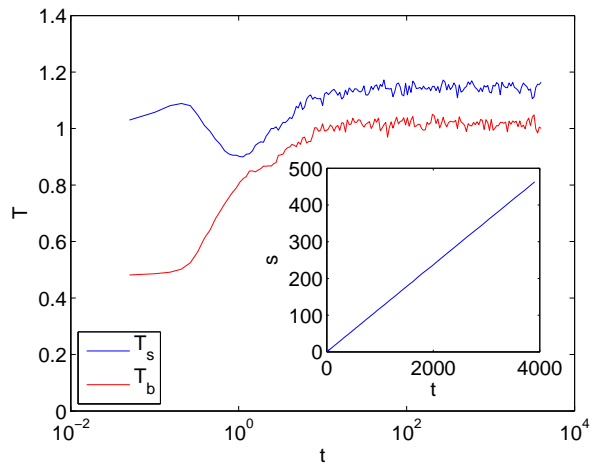}
    \caption{(Color online) Time dependence of the temperature of the small particles ($T_{s}$, upper curve) and
      big particles ($T_{b}$, lower curve) for $\varepsilon=0.90$ and packing
      fraction $\eta=0.5$. Inset: Number of collisions per particle $s$ as a
      function of time in dimensionless units.}
    \label{fig:stationary}
  \end{center}
\end{figure}

In a symmetric binary mixture, we define two temperatures, one for
the big particles ($T_b$) and one for the small ones ($T_s$):
\bea
 T_{b}&=& \frac{m_{b}}{N} \sum_i  \delta_{R_i,R_b} v_i^2\nonumber\\
 T_{s}&=& \frac{m_{s}}{N} \sum_i  \delta_{R_i,R_s} v_i^2.
\label{delta-temp}
\eea 
These are in general unequal in the stationary state due to violation
of equipartition \cite{huthmann}. The analogous cooling equations for
$T_b$ and $T_s$ have been derived, also in the presence of driving,
and can in principle be used to determine the driving amplitude
\cite{kranz-2009}. We refrain from an exact solution and instead
choose the driving amplitude $ p_{Dr} \sim \sqrt{1-\varepsilon^2}$ and
absorb the density dependence in $f_{Dr}$ which is chosen equal to the
collision frequency. An example for the relaxation to the stationary
state is shown in Fig.\ref{fig:stationary}, where we plot the two
temperatures $T_{b}$ and $T_{s}$.

In the following we use units of length, mass and time, such that the
radius of the small particles is one, their mass is one and
$\frac{1}{2}\overline{m_i \mathbf v_i^2(t=0)} =1$ (average over all
particles). The natural time step in an event driven simulation is the
number of collisions. In the stationary state the number of collisions
increases linearly in time, as shown in the inset of
Fig. \ref{fig:stationary} so that the two are related by a constant factor.

\subsection{Long-time tails of the Velocity Autocorrelation}

The VACF is defined by
\begin{equation}
 \Phi (t) = \frac{1}{N} \sum_{i=1}^N\left\langle \mathbf v_i (t) \mathbf v_i (0)\right\rangle,
\end{equation}
where the average $\left\langle \ldots \right\rangle$ is taken over
initial conditions. The VACF is known to exhibit a hydrodynamic long-time tail in systems
of hard spheres \cite{alder, williams-2006} and hard disks
\cite{isobe}. Such long-time tails have also been observed in driven
granular fluids in three dimensions. It was furthermore shown by
simulations and a simple scaling argument that the tails depend
sensitively on the driving mechanism: If driving conserves momentum
locally, then the $t^{-3/2}$-tail known from elastic systems
prevails, whereas it is replaced by $t^{-1}$, if momentum is not
conserved by the driving \cite{fiege}. 

For two-dimensional fluids the situation is less clear. Two
functional forms of the long-time decay of the equilibrium VACF have
been proposed. Hydrodynamic theory \cite{hansen-mcdonald} as well as
mode-coupling theory \cite{ernst} predict $\Phi_{\text{HD}} \propto
t^{-1}$. If these approaches are made self-consistent
\cite{wainwright,kawasaki}, the time decay is changed to
$\Phi_{\text{SC}} \propto t^{-1}(\sqrt{\ln(t)})^{-1}$ in agreement
with renormalization group theory \cite{forster}. Recent simulations
\cite{isobe} claim that the self consistent theory is supported by
large scale simulations.

\begin{figure}[t]
\begin{center}
 \includegraphics[width=\columnwidth]{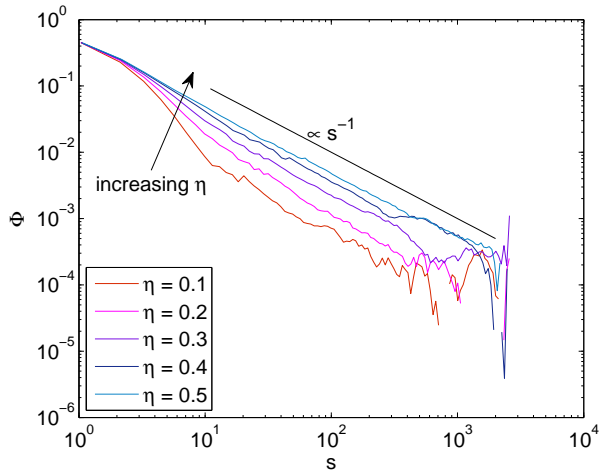}
 \caption{(Color online) The VACF for small particles for low and intermediate packing fractions as a
   function of time measured in number of collisions per particle
   $s$. The long-time tails are more pronounced for increasing
   $\eta$.}
    \label{fig:VACF70}
  \end{center}
\end{figure}

\begin{figure}[t]
\begin{center}
 \includegraphics[width=\columnwidth]{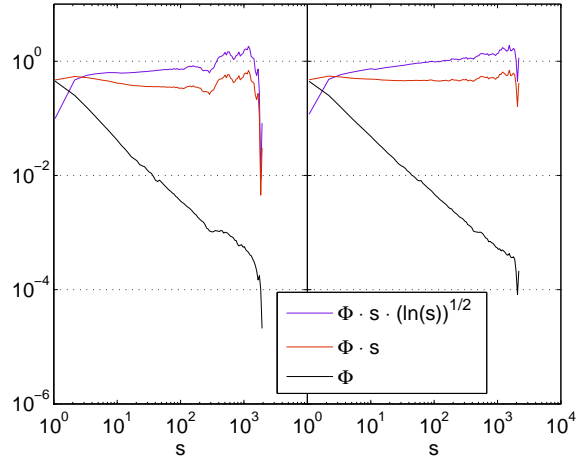}
 \caption{(Color online) The VACF (lower curves) for small particles for $\eta=0.4$ (left) and $\eta= 0.5$ (right) and
   $\varepsilon =0.7$, divided by both predictions $\Phi \propto
   s^{-1}$ (middle), $\propto s^{-1}(\sqrt{\ln(s)})^{-1}$ (upper curves); time given in
   number of collisions per particle $s$.}
    \label{fig:VACF5070}
  \end{center}
\end{figure}

We compute the VACF for a bidisperse mixture of $N= 4\cdot 10^6$
particles as suggested by \cite{isobe} in order to be able to analyze
the long-time behavior of the VACF. The more inelastic system,
$\varepsilon=0.7$, is expected to exhibit the most prominent long-time
tail as observed in three-dimensional inelastic fluids
\cite{fiege}. Fig. \ref{fig:VACF70} shows the VACF for an inelastic
mixture with $\varepsilon = 0.7$ and packing fractions up to
$\eta=0.5$. The long-time tail in the decay seems to be most
pronounced for $\eta=0.4$ and $\eta=0.5$. In order to test the above
predictions, we divide the VACF by each of the suggested decay laws
and show the result in Fig. \ref{fig:VACF5070}. If one of the two
decay laws described the decay exactly, then the resulting curve
should fluctuate around a constant. This is not the case neither for
$\propto t^{-1}$ nor for $\propto t^{-1}\sqrt{\ln(t)}^{-1}$. Hence,
neither of the suggestions is supported by our data, however we cannot
exclude that one has to go to even longer times to identify the
long-time tail unambiguously.

\begin{figure}[t]
\begin{center}
 \includegraphics[width=\columnwidth]{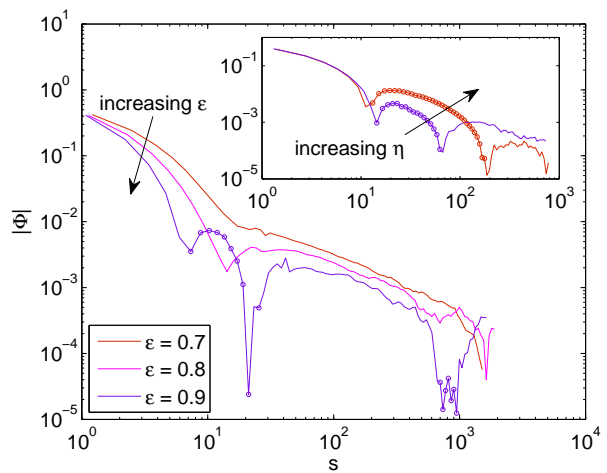}
    \caption{(Color online) The VACF for small particles for different $\varepsilon$ at packing fraction
      $\eta = 0.72$. Marked data points indicate a negative
      value. Inset: The VACF for $\varepsilon = 0.7$ and packing
      fractions $\eta = 0.76, 0.78$; time given in
   number of collisions per particle $s$.}
    \label{fig:VACF_Backscattering}
  \end{center}
\end{figure}

For increasing $\varepsilon$ and high packing fractions the VACF is
expected to show backscattering effects, \ie, a time interval, in
which the VACF becomes negative because the particles are reflected
from the cage formed by their neighbours. This is confirmed by the
VACF in a system of $\eta = 0.72$ where the negative part of the VACF
sets in for larger $\varepsilon$ after showing a dip for intermediate
$\varepsilon$, see Fig. \ref{fig:VACF_Backscattering}. For constant
$\varepsilon$, the range of the negative VACF is increased for
increasing $\eta$, see inset of Fig. \ref{fig:VACF_Backscattering}.

\subsection{Mean Square Displacement and Diffusion Coefficients}

One prominent signature of the glass transition is the formation of a
plateau in the MSD
\be \langle (\Delta
\mathbf{r}(t))^2\rangle=\frac{1}{N}\sum_i^N\langle (
\mathbf{r}_i(t)-\mathbf{r}_i(0))^2\rangle .
\ee
\begin{figure}[t]
  \begin{center}

    \includegraphics[width=\columnwidth]{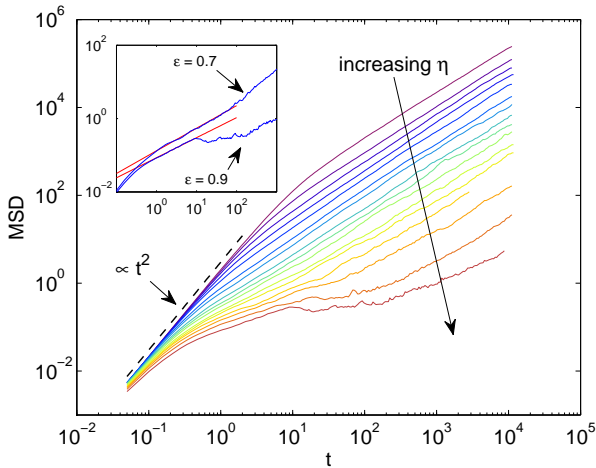}
    \caption{(Color online) MSD for small particles for $\varepsilon=0.90$ and different packing fractions $\eta=0.1$; $0.2$; $0.3$; $0.4$; $0.5$; $0.6$; $0.65$; $0.7$; $0.72$; $0.74$; $0.75$; $0.76$; $0.77$; $0.78$; $0.79$; $0.80$; $0.81$.}
    \label{fig:MSD-eps=90}
  \end{center}
\end{figure}

The MSD for a mildly inelastic system (coefficient of restitution
$\varepsilon=0.9$) is shown in Fig. \ref{fig:MSD-eps=90} for a wide
range of packing fractions $0.1\leq\eta\leq 0.81$. The simulated
system consists of $10000$ particles, which have been simulated for
$10^7$ collisions per particle. Subsequently, the data was divided
into approximately 1000 time intervals each of them giving rise to a
single MSD. These were averaged to obtain the final time dependent MSD
\cite{allen-tildesley} with a standard error of $3\cdot 10^{-4}$.
Ballistic motion is observed for small times for all densities. The
larger the density the smaller the ballistic regime. For packing
fractions $\eta<0.7$, ballistic motion crosses over to diffusive
behaviour at long times. For $\eta\geq 0.7$, a plateau develops,
indicating that a cage has formed which restricts the particles´
motion to the space within the cage. Actually, our data indicate
subdiffusive behavior instead of a plateau; a blowup of the relevant
time range is shown in the inset of Fig. \ref{fig:MSD-eps=90}. For
times $1<t<10$ the MSDs for $\eta=0.81$ show algebraic increase with
exponents $0.6163(11)$ for $\varepsilon=0.7$ and $0.547(6)$ for
$\varepsilon=0.9$, respectively.

For area fraction $0.7\leq \eta
\leq 0.77$ the cage is observed to break up for longer times so that
ultimately diffusion prevails except for the highest density.

\begin{figure}[t]
  \begin{center}

    \includegraphics[width=\columnwidth]{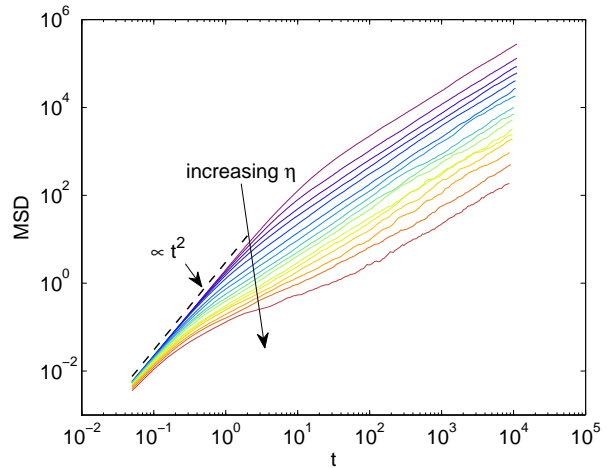}
    \caption{(Color online) MSD for small particles for $\varepsilon=0.70$ and different packing fractions $\eta=0.1$; $0.2$; $0.3$; $0.4$; $0.5$; $0.6$; $0.65$; $0.7$; $0.72$; $0.74$; $0.75$; $0.76$; $0.77$; $0.78$; $0.79$; $0.80$; $0.81$.}
    \label{fig:MSD-eps=70}
  \end{center}
\end{figure}

As the system becomes more inelastic we expect the glass transition to
shift to higher densities \cite{kranz-2010}. Furthermore the range of
densities where a clear separation of timescales can be observed, is
expected to shrink. These expectations are indeed born out by our
simulations. The MSD for a more inelastic system (coefficient of
restitution $\varepsilon=0.7$) is shown in
Fig. \ref{fig:MSD-eps=70}. The plateau is hardly visible as compared
to the more elastic system, instead 
subdiffusive behaviour is observed for the highest densities.

\begin{figure}[t]
  \begin{center}
    \includegraphics[width=\columnwidth]{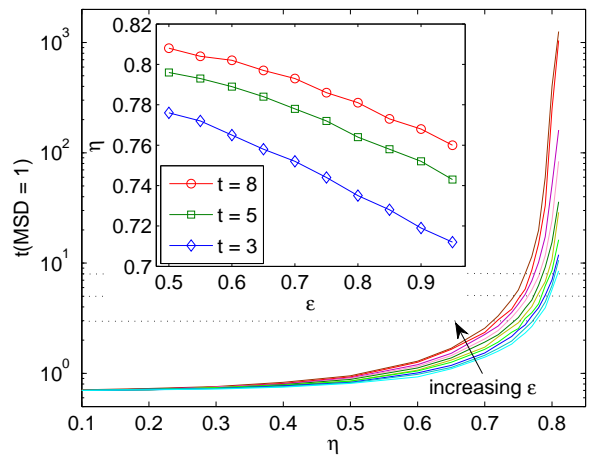}
    \caption{(Color online) Times $t(\eta, \varepsilon)$ of equal MSD for small particles for
      $\varepsilon=0.5$, $0.55$, $0.6$, $0.65$, $0.7$, $0.75$, $0.8$,
      $0.85$, $0.9$, $0.95$. Inset: Lines of equal MSDs for different
      times.}
    \label{fig:MSD_equal}
  \end{center}
\end{figure}

Next, we want to quantify the prediction of granular mode-coupling
theory (GMCT) that
the more inelastic systems require increasing critical densities for
the glass transition to occur. This is done in two steps. We first identify a
timescale, such that the small particles' MSD $=1$. This corresponds to
the time at which the mean travelled distance of all small particles
equals their radius. This timescale, i.e. $t(\text{MSD}=1)$, is shown in
Fig. \ref{fig:MSD_equal} as a function of area fraction for a wide
range of $\varepsilon$. As can be seen in the figure, this time scale
strongly increases as the glass transition is approached, the more so
the more elastic the system. Now we ask: at what area fraction has the
timescale acquired a given value, e.g. $t(\text{MSD}=1)=3$? This yields an
area fraction as a function of $\varepsilon$, which is shown as the
blue curve in the inset and corresponds to the lowest broken line in
the main part of the figure. Hence, in the inset we show lines of
{\it equal relaxation time} ($t=3,5,8$) in the $(\eta, \varepsilon)$ plane.
For all times $t$ the packing fraction $\eta$ is a monotonically
decreasing function of $\varepsilon$ in agreement with GMCT
\cite{kranz-2010}.

The MSD is related to the VACF by $\frac{\partial}{\partial t} \Delta
\mathbf r^2(t) = 2\int_0^t \Phi(t^\prime)\mathrm dt^\prime $. With
this relation we can define a time dependent diffusion coefficient
$D(t) := 1/2 \int_0^t \Phi (t^\prime)\mathrm dt^\prime$. The predicted
functional forms of the VACF, $\Phi_{\text{HD}}(t)\propto t^{-1}$ and
$\Phi_{\text{SC}}(t)\propto t^{-1}(\sqrt{\ln(t)})^{-1}$, respectively,
yield the following time dependent diffusion coefficients:
\begin{eqnarray}
D_{\text{HD}}(t) & \propto &  \ln t ,\label{equ:timedependentD1}\\
D_{\text{SC}} (t) & \propto &\sqrt{\ln t}.
\label{equ:timedependentD}
\end{eqnarray}
Both of these expressions diverge for $ t \rightarrow \infty$, hence for both the long-time limit of the diffusion coefficient does not exist. 

\begin{figure}[t]
  \begin{center}
    \includegraphics[width=\columnwidth]{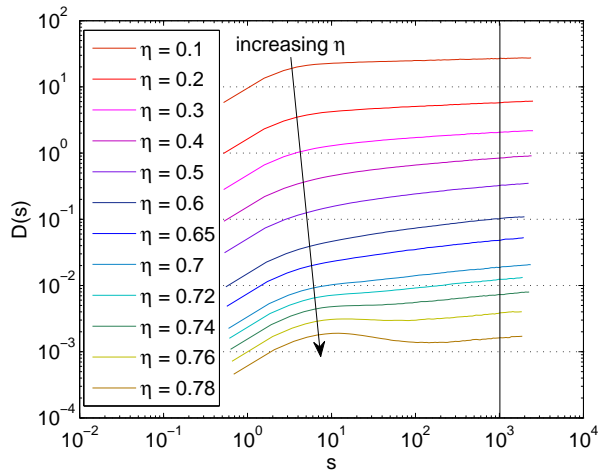}
    \caption{(Color online) Time dependent diffusion coefficient $D(s)$ for small particles for $\varepsilon = 0.7$ and different packing fractions.}
    \label{fig:Dt}
  \end{center}
\end{figure}

We first check that $D(s) = \frac{1}{4} \frac{\partial }{\partial s} \langle \Delta \mathbf{r}(s))^2 \rangle$ does not attain a stationary value but increases indefinitely as a function of time. Data for $\varepsilon =0.7$ are shown in Fig. \ref{fig:Dt} for a wide range of packing fractions. Obviously no stationary value is attained for any packing fraction; the same holds for other values of $\varepsilon$. 

\begin{figure}[tbp]
  \begin{center}
    \includegraphics[width=\columnwidth]{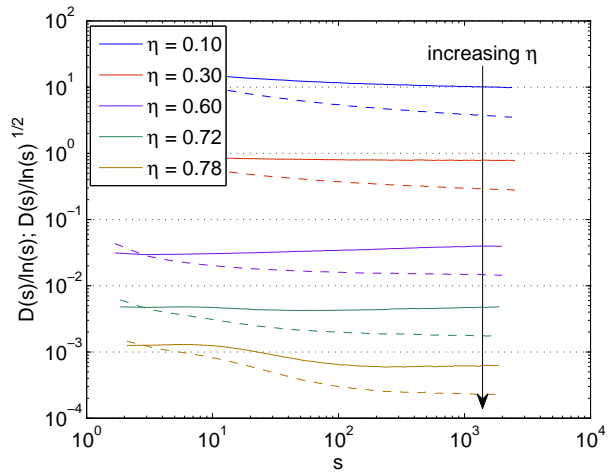}
    \caption{(Color online) Full lines: $D(s)/\ln(s)$, dashed lines: $D(s)/\sqrt{\ln(s)}$.}
    \label{fig:Dtcheck}
  \end{center}
\end{figure}

In order to check if one of the proposed time dependences in Eqs. (\ref{equ:timedependentD1}) or (\ref{equ:timedependentD}) is correct, we plot $D(s)/\ln(s)$ and $D(s)/\sqrt{\ln(s)}$. Similar to the test of the VACF, these quantities should fluctuate around a constant if they were describing $D(s)$ correctly. The resulting graphs are plotted in Fig. \ref{fig:Dtcheck}.

Neither the prediction of MCT, $D(t)\propto \sqrt{\ln(t)}$, nor the
prediction of hydrodynamics, $D(t) \propto \ln (t)$, account for the
data in the full range of densities. MCT works slightly better at
higher densities, whereas hydrodynamics fits the data better at
intermediate densities. Hence, our data are not sufficient to
discriminate between the two alternatives.

\begin{figure}[t]
  \begin{center}
    \includegraphics[width=\columnwidth]{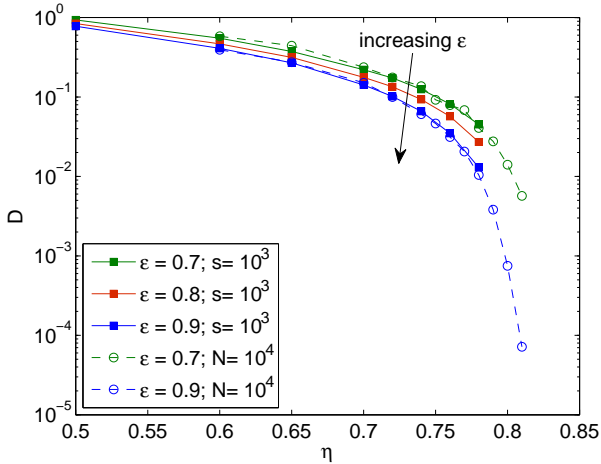}
    \caption{(Color online) Diffusion coefficients for small particles extracted from the different data sets; for detailed explanation see text.}
    \label{fig:D_eta}
  \end{center}
\end{figure}

The above discussion shows that the long-time behavior in a
two-dimensional fluid is not truely diffusive. On the other hand it is
known that the long-time tails of the VACF are suppressed close to the
glass transition, where the shear viscosity $\nu$ increases
dramatically. In fact, the MCT of Ernst {\it et al.} \cite{ernst}
predicts for a two-dimensional elastic fluid
\begin{equation}
 \Phi(t) \propto \frac{kT}{8m\pi (\nu+D)} \frac{1}{t}.
\end{equation}
For large packing fractions in the limit of the glass transition,
$\nu$ diverges and thus, the long-time tail in the VACF is
suppressed. This allows us to calculate the diffusion coefficient
approximately by two methods:
\begin{itemize}
\item We calculate the time dependent diffusion coefficient and
  estimate the diffusion coefficient by its value when the mean number
  of collisions per particle is $s=1000$, \ie, $D=D(s=1000) $.
\item We use a smaller system of $N=10^4$ particles where the
  long-time tails are cut off due to finite size effects. Hence the
  long-time limit of the integral over the VACF is constant and allows
  to extract the diffusion constant \cite{FH_private_comm}.
\end{itemize}
The diffusion coefficients extracted by these two different methods
are depicted in Fig. \ref{fig:D_eta}. First, we note that the results
of both methods agree quantitatively, so that it is indeed possible to
estimate the diffusion coefficient from our data. Second, the
diffusion coefficient decreases by several orders of magnitude as the
glass transition is approached. Third, for more inelastic systems the
decrease of $D$ is shifted to higher densities.

\subsection{Incoherent structure function}

Complete information about the motion of a tagged particle is
contained in the incoherent scattering function, defined as
 \be
S_{\text{inc}}(k,t)=\frac{1}{N}\sum_{i=1}^N \left\langle e^{\mathrm i
    \mathbf{k}\cdot(\mathbf{r}_i(t)-\mathbf{r}_i(0))}\right\rangle.
\ee
In a molecular as well as in a colloidal glass,
this function displays a two-step decay: The so called
$\beta$-relaxation towards a plateau and the subsequent
$\alpha$-relaxation for asymptotically long times. The timescale of
the $\alpha$-relaxation diverges as the glass transition is
approached. 

\begin{figure}[t] 
  \begin{center}
    \includegraphics[width=\columnwidth]{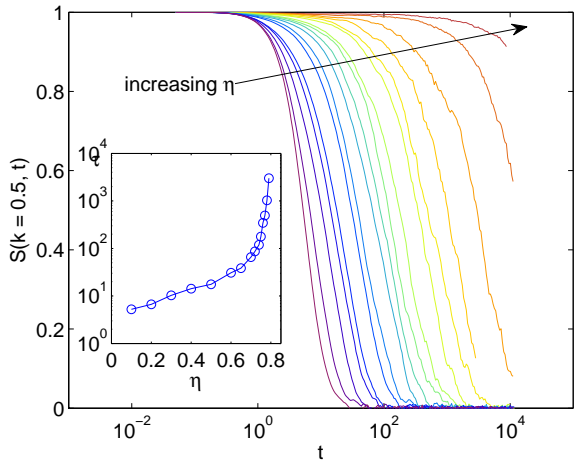}
    \caption{(Color online) $S_{\text{inc}}(k=0.5,t)$ for small particles for $\varepsilon=0.90$ and different packing fractions $\eta=0.1$; $0.2$; $0.3$; $0.4$; $0.5$; $0.6$; $0.65$; $0.7$; $0.72$; $0.74$; $0.75$; $0.76$; $0.77$; $0.78$; $0.79$; $0.80$; $0.81$. Inset: Relaxation time $\tau$ at which the incoherent scattering function has decayed to half its initial value as a function of packing fraction.}
    \label{fig:Sq05t}
  \end{center}
\end{figure}

In Fig. \ref{fig:Sq05t}, we plot the
incoherent scattering function for $\varepsilon=0.9$ and increasing
packing fraction up to $\eta=0.81$. One clearly observes a pronounced
slowing down which is quantified in the inset of
Fig. \ref{fig:Sq05t}. There we plot the relaxation time
$\tau(\eta)$, the time at which $S_{\text{inc}}(k=0.5,t)$ has decayed to half its
initial value. This relaxation time is seen to diverge as the glass
transition is approached. 

\begin{figure}[t]
  \begin{center}

    \includegraphics[width=\columnwidth]{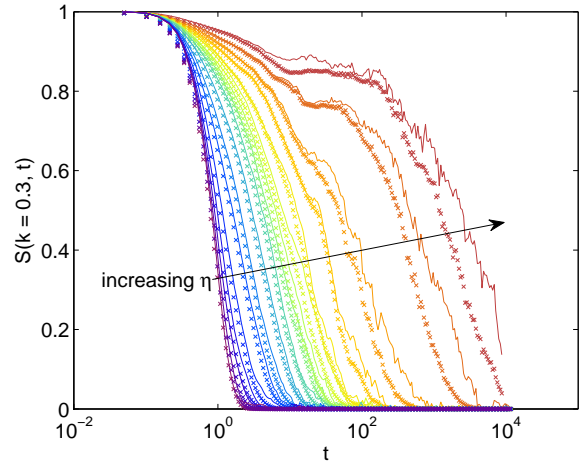}
    \caption{(Color online) $S_{\text{inc}}(k=3,t)$ (full lines) for small particles in comparison to the
      Gaussian approximation (crosses); parameters as in Fig. \ref{fig:Sq05t}}.
    \label{fig:Sq3t}
  \end{center}
\end{figure}

The two-step relaxation is related to the plateau in the MSD and hence
more easily observed for larger wave numbers. We plot in
Fig. \ref{fig:Sq3t} the incoherent scattering functions
$S_{\text{inc}}(k=3,t)$ for various densities; a two-step relaxation
is clearly observed for packing fractions $\eta \geq 0.78$. We also compare
$S_{\text{inc}}(k,t)$ to the Gaussian approximation
 \begin{equation}
S^{\text{Gauss}}_{\text{inc}}(k,t)= e^{-k^2\langle (\Delta
\mathbf{r}(t))^2\rangle/4}. 
\end{equation}
As expected the Gaussian approximation works well for small densities
at all times, whereas for the largest densities only the short time
behaviour is approximately described by a Gaussian. The data for the
incoherent scattering function look rather similar to those of the
three-dimensional system which have been discussed in detail in
\cite{kvl-2011}.

\subsection{Dynamic heterogeneity}

\begin{figure}[t]
  \begin{center}
    \includegraphics[width=\columnwidth]{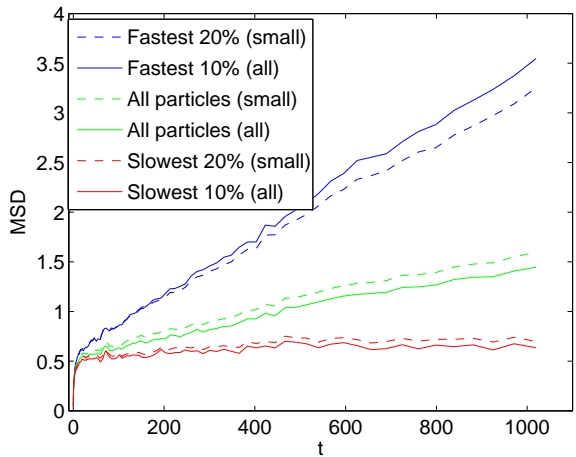}
    \caption{(Color online) MSD for fast and slow particles, either restricted to small particles only (dashed) or for all particles (full line); the two upper curves refer to the fast particles, the two lower curves to the slow ones, and the two middle curves to all particles. The MSD at $t=10^3$ has been used to identify slow and fast particles.}
    \label{fig:Dyn_Het}
  \end{center}
\end{figure}

We expect the dynamics to become increasingly heterogeneous as the
glass transition is approached. To demonstrate this point, we plot in
Fig. \ref{fig:Dyn_Het} the MSD of the $10\%$ slowest and $10\%$
fastest particles as compared to an average over all particles in a
system of packing fraction $\eta = 0.8$ and $\varepsilon=0.9$. Whereas
the slow particles are completely immobile, the fast ones are seen to
move diffusively covering distances which correspond to several
radii. We plan to analyze dynamic heterogeneities in a future project.

\subsection{Pair correlation and structure factor}

In this section we want to analyze the structure of a dense granular
fluid and in particular exclude long range crystalline
order. Translational order is detected by
the pair correlation function, $g(r)$, which gives the probability to
find a particle in a given distance $r$, from 
another particle at the origin:
\be
g(r) = \frac{1}{n_0 N} \left\langle \sum_{j\ne i}
  \delta (\mathbf{r}-\mathbf{r}_{ij}) \right\rangle,
\ee
where $n_0$ is number density. For a homogeneous system, $g(r)$ approaches a constant as
$r\to\infty$ which is unity with the chosen normalization.
Practically, $g(r)$ is calculated as follows:
\be
g(r) = \frac{1}{2\pi r \Delta r n_0 N} \left\langle \sum_{j\ne i}
  \theta(r+\Delta r -r_{ij})  \theta(r_{ij}-r)\right\rangle.
\ee
The bin size $\Delta r=L/800$ has been chosen such that the number of
bins and the average number of particles per bin remain constant for all packing fractions. This corresponds to
$\Delta r=0.83$ for the most dilute ($\eta=0.1$) and 
$\Delta r=0.30$ for the densest ($\eta=0.81$) system. 

\begin{figure}[t]
  \begin{center}
    \includegraphics[width=\columnwidth]{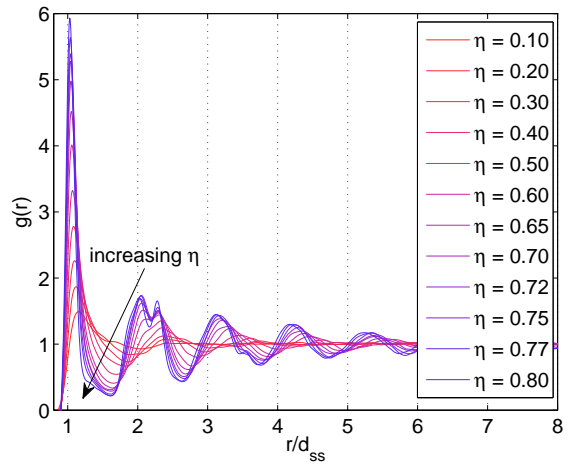}
    \caption{(Color online) Pair correlation function for small particles
      for $\varepsilon=0.90$ and different packing fractions, where
      $d_{ss}$ indicates the sum of two small particles' radii.}
    \label{fig:PCF-1-eps=90}
  \end{center}
\end{figure}

\begin{figure}
  \begin{center}
    \includegraphics[width=1.0\columnwidth]{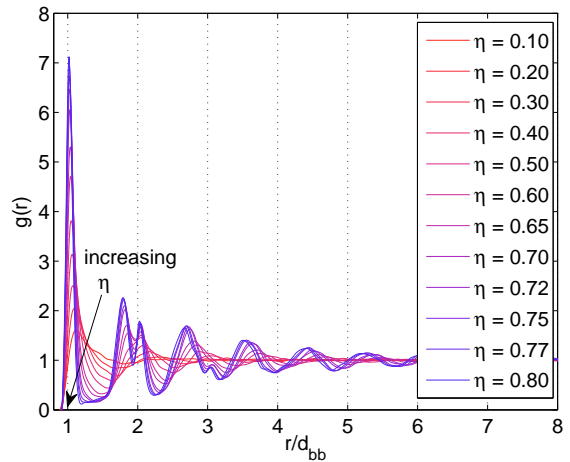}
    \caption{(Color online) Pair correlation function for big particles for
      $\varepsilon=0.90$ and different packing fractions, where
      $d_{bb}$ indicates the sum of two big particles' radii.}
    \label{fig:PCF-2-eps=90}
  \end{center}
\end{figure}

For binary mixtures with big and small particles, there are three
different pair correlation functions: between big particles only,
between big and small particles, and between small particles only.
All three radial distribution functions vanish for $r/d_{ij}<1$,
because the particles are hard, and have a global peak at the contact
point, $r/d_{ij}=1$. In Figs. \ref{fig:PCF-1-eps=90} and
\ref{fig:PCF-2-eps=90} we show pair correlations for small and big
particles with size ratio $1.25$ as a function of packing
fraction. The data have been smoothed, implying a continuous growth of
the correlations for small distances, whereas the raw data strictly
vanish for $r/d_{ij} < 1$. Increasing the packing fraction, the peaks
in $g(r)$ get higher and sharper. Simultaneously oscillations develop
for larger $r$. At packing fractions around $\eta\sim 0.55$, the
second peak splits into two separate peaks, a phenomenon that is more
pronounced for the big particles. This splitting of the second peak
near $r/d_{ij}=2$ is a universal phenomena in glasses, but not a good
indicator of the transition itself. Our results are in good agreement
with the experimental ones of ref. \cite{abate-2006}.


\begin{figure}[t]
  \begin{center}
    \includegraphics[width=\columnwidth]{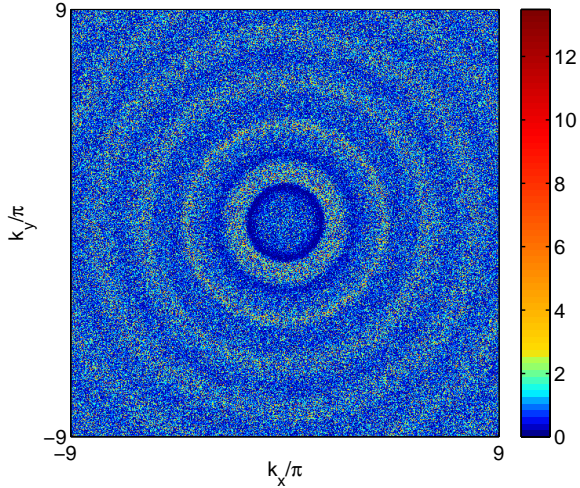}
    \caption{(Color online) Structure factor for $\varepsilon=0.90$ and
      $\eta=0.81$, for small particles. The contour plot of the
      structure factor shows a circular pattern which is
      characteristic of an amorphous system.}
    \label{fig:SF-eps=90}
  \end{center}
\end{figure}

We have chosen to study a bidisperse system in order to avoid
crystallization. Hence we need to verify that our system does indeed
remain structurally disordered. We do so by mimicking a scattering
experiment: We compute the static structure factor
\be
S(\mathbf k) = \left\langle \rho_{\mathbf k}(t) \rho_{-\mathbf k}(t)\right\rangle,
\ee
where $\mathbf k$ is the wave vector and $\rho(\mathbf k)$ is the collective
density variable, defined as
\be
\rho_{\mathbf k}(t) = \frac{1}{\sqrt{N}} \sum_{j=1}^N \mbox{e}^{\mathrm i\mathbf k \cdot \mathbf r_j(t)}.
\ee
The computed structure factor is shown in Fig. \ref{fig:SF-eps=90}
for the case of $\varepsilon=0.90$ and $\eta=0.80$. The pattern is
clearly isotropic, revealing the prefered distances which are visible
as peaks in $g(r)$. We conclude that crytallization has been avoided
and our system remains structurally disordered even when the particles
-- or a large fraction of them -- are arrested.

An alternative measure ($Q_6$) for orientational order was introduced
in ref. \cite{Steinhardt} and applied to granular media in 
ref. \cite{Torquato}. We have computed $Q_6$ for a range of packing
fractions and find no indication of long range orientational order in
agreement with ref. \cite{Torquato}.
 
\section{Conclusion and outlook}

Simulations of a two-dimensional granular fluid in a stationary state
reveal strong signatures of a glass transition, as the packing fraction
is increased to approximately $\eta=0.8$. The relaxation time of the
incoherent scattering function diverges; the MSD
develops a plateau; the diffusion constant goes to zero; the dynamics
becomes increasingly heterogeneous. The results of event driven
simulations are in qualitative agreement with recent experiments using vibro-fluidized \cite{reis-2007} and air
fluidized \cite{abate-2006} granular systems.

We not only confirm the experimental results, but in addition
investigate, how glassy dynamics depends on the degree of energy
dissipation. In particular, we have tested the predictions of a recent
mode-coupling theory \cite{kranz-2010} for the dependence of the glass transition on
inelasticity. The theory predicts that the transition is pushed to
higher densities the more inelastic the system is. To check this
point, we have simulated a range of coefficients of restitution,
$0.5\leq \varepsilon \leq 0.9$, and compared the mean square displacements
as a function of $\varepsilon$. The time $\tau(\varepsilon,\eta)$,
where $\langle \Delta \mathbf r^2(\tau) \rangle=1$ is shifted to higher densities -- in agreement with
the theory. Similar behaviour is found for the lines of equal
diffusivities. In addition to the shift in density, mode-coupling
theory predicts that the plateau in the MSD is less pronounced for the
more inelastic system. This prediction is also born out by our
simulations.

Our simulations clearly show that the diffusion coefficient does not exist in a two-dimensional granular fluid; an appropriately defined time dependent diffusion coefficient grows indefinitely for long times. However, we are not able to extract the functional form unambiguously. The singular contribution is suppressed near the glass transition due to the divergence of the static shear viscosity, so that our estimates of the diffusion coefficient become increasingly more reliable as the glass transition is approached.

Dynamic heterogeneities have been investigated extensively in molecular glasses \cite{berthier}. Various tools, such as multipoint correlations, have been used to detect increasing spatial correlations in a disordered structure. For the future, we plan to use these methods in order to analyze dynamic heterogeneities in a granular medium.

\section{Acknowledgements}

We thank Elmar Staerk and Matthias Sperl for providing us with initial
configurations and Matthias Sperl, Timo Aspelmeier, Katharina
Vollmayr-Lee, Felix H\"ofling and Till Kranz for many useful discussions. We furthermore acknowledge support from the DFG by FOR 1394.


\end{document}